# Electron heating in the laser and static electric and magnetic fields

Yanzeng Zhang, and S. I. Krasheninnikov

# Electron heating in the laser and static electric and magnetic fields

Yanzeng Zhang and S. I. Krasheninnikov
*University of California San Diego, 9500 Gilman Dr., La Jolla, California 92095, USA*

A 2D slab approximation of the interactions of electrons with intense linearly polarized laser radiation and static electric and magnetic fields is widely used for both numerical simulations and simplified semi-analytical models. It is shown that in this case, electron dynamics can be conveniently described in the framework of the 3/2 dimensional Hamiltonian approach. The electron acceleration beyond a standard ponderomotive scaling, caused by the synergistic effects of the laser and static electro-magnetic fields, is due to an onset of stochastic electron motion. *Published by AIP Publishing.* https://doi.org/10.1063/1.5016976

The generation of high-energy electron beams in the course of the interactions of an intense laser with plasma is of great interest for many different applications (e.g., ion acceleration, X-ray generation, positron production, etc.). In the past two decades, many results of theoretical studies and simulations (e.g., see Refs. 1–13 and the references therein) have suggested that the synergistic effects of intense laser radiation and static electric and magnetic fields could significantly increase both the energy of the beam electrons and the efficiency of laser-electron coupling. Furthermore, the available experimental data (e.g., see Refs. 13–17 and the references therein) also support these conclusions.

However, the mechanism(-s) of such synergistic effects is still under debate. In Ref. 3, the synergy of the linearly polarized laser radiation propagating in the z-direction with only the y-component of the vector potential, static electric field (in the y-direction) and magnetic field (in the x-direction), was attributed to betatron resonance. However, in Ref. 8, it was shown that the synergy persists for arbitrary orientation of the laser vector potential and the static electric field, while the synergistic effect of the laser and static electric field is due to the "parametric amplification". In Ref. 6, it was shown that the synergistic effects, causing electron heating beyond the ponderomotive scaling,[18] are also present in the case where electrostatic potential, U, depends on the z-coordinate (the direction of laser beam propagation). In Ref. 7, it was demonstrated that in the case of a V-shape electrostatic potential, $U(z) = E_0|z|$, the electron dynamics can be described by a Chirikov-like map[19–21] and a strong electron heating is due to an onset of stochasticity which is determined by a particular relationship between normalized laser vector potential and $E_0$. Later on, the synergy between laser radiation and electrostatic potential U(z) was also reported in Ref. 9. In Ref. 2, it was shown that an onset of stochastic electron motion can also be triggered by the synergy of laser radiation and the constant magnetic field perpendicular to the laser propagation direction. As we can see, there is still no clear vision of the mechanism(-s) of enhanced electron heating due to the synergy of intense laser radiation and static electric and magnetic fields.

In Ref. 22, it was shown that electron dynamics in homogeneous magnetic and linearly polarized laser fields can be described within the simple 3/2 dimensional Hamiltonian framework. In this letter, we show that electron interactions with intense linearly polarized laser radiation and very general static electric and magnetic fields in a 2D slab approximation (including those considered in Refs. 1–12) can also be conveniently described in the framework of the 3/2 dimensional Hamiltonian approach. It allows us to utilize the fundamental results of previous studies on regular and stochastic motion in Hamiltonian systems (e.g., see Refs. 17–19 and the references therein).

To describe electron dynamics in the laser and static electric and magnetic fields, we introduce dimensionless time $\hat{t} = t\omega$, coordinates $\hat{\vec{r}} = k\vec{r}$, velocity $\hat{\vec{v}} = \vec{v}/c$, and vector potential $\hat{\vec{A}} = e\vec{A}/mc^2$, where e is the elementary charge, m is the electron mass, c is the speed of light, and $\omega$ and k ($\omega = kc$) are the frequency and wave number of the laser radiation, respectively. In the rest of this letter, we will omit "hats" over our dimensionless quantities to simplify our expressions. We will also assume that the laser electromagnetic field, which is determined by the vector potential $\vec{A}$, is described in a plane-wave approximation, $\vec{A}(t, \vec{r}) = \vec{A}(t - z)$.

In these dimensionless units, the relativistic equations for the electron motion and $\gamma$-factor can be written as follows:

$$\frac{dP_\alpha}{dt} = \frac{dA_\alpha}{dt} - \frac{\partial A_\beta}{\partial x_\alpha} v_\beta, \qquad (1)$$

$$\frac{d\gamma}{dt} = \frac{\partial A_\beta}{\partial t} v_\beta, \qquad (2)$$

where

$$P_\alpha = \gamma dx_\alpha/dt \equiv \gamma v_\alpha \qquad (3)$$

and

$$\gamma^2 = 1 + \vec{P}^2. \qquad (4)$$

Then, we consider different cases of the orientations of both the laser field and the static electric and magnetic fields (see Fig. 1). Some of them were the subjects of previous studies.[1–12]

*Case 1.* We start with the case where both laser and static electric fields are in the y-direction, while the static magnetic field (depending only on the y-coordinate) is in the x-direction (similar to that considered in Refs. 2, 8, and 11). Thus, we have

$$\vec{A} = \vec{e}_y\{\tilde{A}_y(t-z) - t\partial U(y)/\partial y\} + \vec{e}_z A_B(y), \quad (5)$$

where $\vec{e}_{(\ldots)}$ are the unit vectors, while $U(y)$ and $A_B(y)$ describe the electrostatic potential and the x-component of the magnetic field. Then, from Eqs. (1) and (2), we have

$$\frac{dP_x}{dt} = 0 \rightarrow P_x = \bar{P}_x = P_x|_{t=0}, \quad (6)$$

$$\frac{d(P_y - \tilde{A}_y)}{dt} = -\frac{\partial U(y)}{\partial y} - \frac{\partial A_B(y)}{\partial y}v_z, \quad (7)$$

$$\frac{d(P_z - A_B)}{dt} = -\frac{\partial \tilde{A}_y}{\partial z}v_y, \quad (8)$$

$$\frac{d(\gamma + U)}{dt} = \frac{\partial \tilde{A}_y}{\partial t}v_y. \quad (9)$$

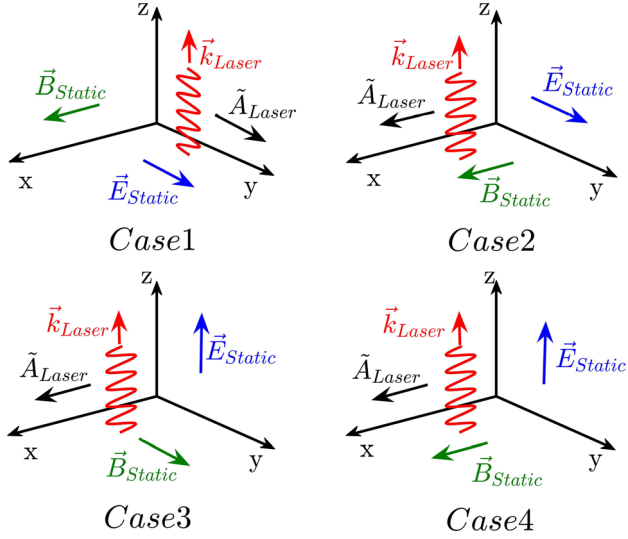

FIG. 1. The sketches of different orientations of both the laser field and the static electric and magnetic fields for four cases.

Combining Eqs. (8) and (9), we find

$$\gamma - P_z + W^{(+)}(y) = C_\perp \equiv \{\gamma - P_z + W^{(+)}(y)\}_{t=0}, \quad (10)$$

where $W^{(+)}(y) = U(y) + A_B(y)$. From Eqs. (4) and (10), we find

$$\gamma = \frac{1}{2}\left\{\frac{1 + \bar{P}_x^2 + P_y^2}{C_\perp - W^{(+)}(y)} + C_\perp - W^{(+)}(y)\right\}. \quad (11)$$

Introducing the variable $\xi = t - z$ and using Eq. (10), we obtain

$$\frac{d\xi}{dt} = \frac{\gamma - P_z}{\gamma} = \frac{C_\perp - W^{(+)}(y)}{\gamma}, \quad (12)$$

while from Eqs. (3) and (7), we find

$$\frac{C_\perp - W^{(+)}(y)}{\gamma}\frac{d(P_y - \tilde{A}_y)}{d\xi} = -\frac{\partial U(y)}{\partial y} - \frac{\partial A_B(y)}{\partial y}\frac{P_z}{\gamma}$$
$$\equiv -\frac{\partial W^{(+)}(y)}{\partial y} - \frac{\partial A_B(y)}{\partial y}\frac{W^{(+)}(y) - C_\perp}{\gamma}. \quad (13)$$

Inserting expression (11) into Eq. (13), we have

$$\frac{d\tilde{P}_y}{d\xi} = -\frac{1}{2}\left\{\frac{1 + \bar{P}_x^2 + (\tilde{P}_y + \tilde{A}_y)^2}{[C_\perp - W^{(+)}(y)]^2}\right\}\frac{\partial W^{(+)}(y)}{\partial y}$$
$$-\frac{1}{2}\frac{\partial W^{(-)}(y)}{\partial y}, \quad (14)$$

where $\tilde{P}_y = P_y - \tilde{A}_y$ and $W^{(-)}(y) = U(y) - A_B(y)$. Recalling Eq. (3) and using Eq. (12), we find

$$\frac{dy}{d\xi} = \frac{\tilde{P}_y + \tilde{A}_y}{C_\perp - W^{(+)}(y)}. \quad (15)$$

As a result, we arrive to the Hamiltonian equations

$$\frac{d\tilde{P}_y}{d\xi} = -\frac{\partial H_1}{\partial y}, \quad \frac{dy}{d\xi} = \frac{\partial H_1}{\partial \tilde{P}_y}, \quad (16)$$

where

$$H_1 = \frac{1}{2}\left\{\frac{1 + \bar{P}_x^2 + (\tilde{P}_y + \tilde{A}_y)^2}{C_\perp - W^{(+)}(y)} + W^{(-)}(y)\right\}. \quad (17)$$

*Case 2.* Next, we consider the situation where both the laser vector potential and the static magnetic field (again depending only on the y-coordinate) are in the x-direction, while static electric fields are in the y-direction (similar to that considered in Refs. 3 and 11). As a result, we have

$$\vec{A} = \vec{e}_x\tilde{A}_x(t-z) - \vec{e}_y t\partial U(y)/\partial y + \vec{e}_z A_B(y). \quad (18)$$

Then, from Eqs. (1) and (2), we have

$$P_x = \tilde{A}_x + \tilde{\bar{P}}_x, \quad (19)$$

$$\frac{dP_y}{dt} = -\frac{\partial U(y)}{\partial y} - \frac{\partial A_B(y)}{\partial y}v_z, \quad (20)$$

$$\frac{d(P_z - A_B)}{dt} = -\frac{\partial \tilde{A}_x}{\partial z}v_x, \quad (21)$$

$$\frac{d(\gamma + U)}{dt} = \frac{\partial \tilde{A}_x}{\partial t}v_x, \quad (22)$$

where $\tilde{\bar{P}}_x = (P_x - \tilde{A}_x)|_{t=0}$. From Eqs. (3) and (19)–(22), after some algebra similar to that in *case 1*, we arrive to the following Hamiltonian equations:

$$\frac{dP_y}{d\xi} = -\frac{\partial H_2}{\partial y}, \quad \frac{dy}{d\xi} = \frac{\partial H_2}{\partial P_y}, \quad (23)$$

where

$$H_2 = \frac{1}{2}\left\{\frac{1 + (\tilde{A}_x + \tilde{\bar{P}}_x)^2 + P_y^2}{C_\perp - W^{(+)}(y)} + W^{(-)}(y)\right\}. \quad (24)$$

Next, we consider the cases where the static electric field is in the z-direction (such a situation occurs in pre-plasma, see Refs. 6, 7, 9, and 10). For completeness, we also assume that there is a static magnetic field, which depends on z, either parallel or perpendicular to the laser vector potential.

*Case 3*. We start with the case where the static magnetic field, depending on z, is perpendicular to the laser vector potential. Therefore, we have

$$\vec{A} = \vec{e}_x \tilde{A}_x(t-z) - \vec{e}_z \partial U(z)/\partial z + \vec{e}_x A_B(z). \quad (25)$$

Then, from Eqs. (1)–(3), we find

$$P_x - \tilde{A}_x - A_B = C_x \equiv (P_x - \tilde{A}_x - A_B)_{t=0}, \quad (26)$$

$$P_y = \bar{P}_y = P_y|_{t=0}, \quad (27)$$

$$\frac{dP_z}{dt} = -\frac{\partial U(z)}{\partial z} - \frac{\partial A_B(z)}{\partial z} v_x - \frac{\partial \tilde{A}_x}{\partial z} v_x, \quad (28)$$

$$\frac{d(\gamma + U)}{dt} = \frac{\partial \tilde{A}_x}{\partial t} v_x. \quad (29)$$

We introduce variables $\delta = \gamma - P_z$ and $\xi = t - z$. Then, using the expression for $\delta$ and Eqs. (4), (26), and (27), we find

$$\frac{dz}{d\xi} = \frac{P_z}{\delta}, \quad P_z = \frac{1 + \bar{P}_y^2 + (C_x + \tilde{A}_x + A_B)^2}{2\delta} - \frac{\delta}{2}. \quad (30)$$

From Eqs. (28) and (29), we derive

$$\frac{d\delta}{d\xi} = \frac{1}{2\delta} \frac{\partial}{\partial z} (C_x + \tilde{A}_x + A_B)^2 + \frac{\partial U}{\partial z}. \quad (31)$$

Observing Eqs. (30) and (31), we obtain the following Hamiltonian equations:

$$\frac{d\delta}{d\xi} = \frac{\partial H_3}{\partial z}, \quad \frac{dz}{d\xi} = -\frac{\partial H_3}{\partial \delta}, \quad (32)$$

where

$$H_3 = \frac{1}{2} \left\{ \frac{1 + \bar{P}_y^2 + (C_x + \tilde{A}_x + A_B)^2}{\delta} + \delta \right\} + U(z). \quad (33)$$

*Case 4*. Finally, we consider the case where the static magnetic field, which depends on z, is parallel to the laser vector potential. This gives

$$\vec{A} = \vec{e}_x \tilde{A}_x(t-z) - \vec{e}_z \partial U(z)/\partial z + \vec{e}_y A_B(z). \quad (34)$$

Therefore, from Eqs. (1), (2), and (4), we find

$$P_x = \tilde{A}_x + \tilde{\bar{P}}_x, \quad (35)$$

$$P_y - A_B = C_y \equiv (P_y - A_B)_{t=0}, \quad (36)$$

$$\frac{dP_z}{dt} = -\frac{\partial U(z)}{\partial z} - \frac{\partial A_B(z)}{\partial z} v_y - \frac{\partial \tilde{A}_x}{\partial z} v_x, \quad (37)$$

$$\frac{d(\gamma + U)}{dt} = \frac{\partial \tilde{A}_x}{\partial t} v_x, \quad (38)$$

where $\tilde{\bar{P}}_x = (P_x - \tilde{A}_x)|_{t=0}$. Again, introducing variables $\delta = \gamma - P_z$ and $\xi = t - z$ and recalling that $d\xi/dt = (\gamma - P_z)/\gamma$, from Eqs. (4) and (35)–(38) after some algebra similar to that in *case 3*, we find

$$\frac{d\delta}{d\xi} = \frac{\partial H_4}{\partial z}, \quad \frac{dz}{d\xi} = -\frac{\partial H_4}{\partial \delta}, \quad (39)$$

where

$$H_4 = \frac{1}{2} \left\{ \frac{1 + (\tilde{A}_x + \tilde{\bar{P}}_x)^2 + (C_y + A_B)^2}{\delta} + \delta \right\} + U(z). \quad (40)$$

As we see, all our Hamiltonians depend not only on the spatial dependence of the vector potential by static electric and magnetic fields but also on the initial momenta and coordinates of electrons. Nonetheless, we will see that electron acceleration caused by the synergy of the laser and static electric and magnetic fields can only be due to an onset of stochasticity which is well described in the framework of the 3/2 dimensional Hamiltonian approach.

To illustrate the transition into stochastic electron motion, we perform numerical integrations of the Hamiltonian equations assuming that $\vec{A}(\xi) = \vec{e}_{(\ldots)} a_0 \sin(\xi)$, where $a_0$ is the normalized amplitude of the laser vector potential and $\vec{e}_{(\ldots)}$ determines the direction of laser polarization. In the following, we consider the quadratic dependence of the components of the static part of the vector potential on the coordinate: $A_B(\zeta) = \kappa_B \zeta^2/2$ and $U(\zeta) = \kappa_U \zeta^2/2$, where $\zeta$ is the coordinate (y or z) and $\kappa_U$ and $\kappa_B$ are some constants. We will assume that $\kappa_U > 0$, while $\kappa_B$, generally speaking, can be both positive and negative. As a result, the functions $W^{(+)}(y)$ and $W^{(-)}(y)$ can be both positive and negative. Here, we will consider the case where $\kappa_B > 0$ so that $W^{(+)}(y) \geq 0$. However, in the case where $\kappa_B > \kappa_U$, we have $W^{(-)}(y) \leq 0$. As a result, for this case, the dependence of both $H_1$ and $H_2$ on y can have three extrema, and we can have negative "energies" $E_1 = H_1$ and $E_2 = H_2$. These energies will be preserved for $a_0 = 0$, but for $a_0 > 0$ and an onset of stochasticity, these energies can be accessible even if initial electron energies are positive. In all our numerical simulations and the below results, we take $\kappa_U = 0.3$ and $\kappa_B = 0.5$.

In *Cases 1 and 2*, we start with $\xi = 0$ and take $y(\xi = 0) = 0$, $C_\perp = 8$, and $\bar{P}_x = \tilde{\bar{P}}_x = 0$ while varying initial "energies" $E_1^{(0)} = H_1|_{\xi=0}$ and $E_2^{(0)} = H_2|_{\xi=0}$ by a proper choice of initial $P_y$. In *Cases 3 and 4*, we start with $\xi = 0$ and take $z(\xi = 0) = 0$, $C_x = C_y = 0$, and $\bar{P}_y = \tilde{\bar{P}}_x = 0$ and vary initial "energies" $E_3^{(0)} = H_3|_{\xi=0}$ and $E_4^{(0)} = H_4|_{\xi=0}$ by a proper choice of initial $\delta$. We present the results of our simulations as the Poincaré maps[17–19] on 2D "energy" $E_{(\ldots)}$ and laser phase $\Delta \xi$ ($0 \leq \Delta \xi \leq \pi$) space. The laser phase is defined as $\Delta \xi \equiv \xi_{cr} - n\pi$, where $n \equiv [\xi/\pi]$ and $\xi_{cr}$ corresponds to the value of $\xi$ when the electron is "crossing" some particular boundary. These boundaries correspond to $P_y = 0$ (*Cases 1 and 2*) and $z = 0$ (*Cases 3 and 4*). Our simulation results show that the electron dynamics in *case 2* and *case 4* is rather similar to that in *case 1* and *case 3*, respectively. Therefore, due to the lack of space, we only present the results for *Cases 1* and *3* here.

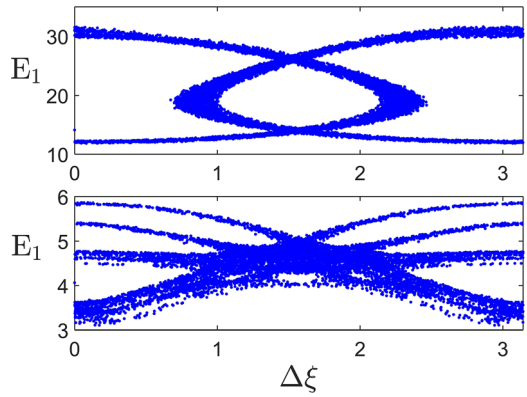

FIG. 2. The "map" obtained from numerical simulation for *case 1*, $a_0 = 0.5$, and initial energy $E_1^{(0)} = 4$ in the bottom panel and $E_1^{(0)} = 14$ in the top panel.

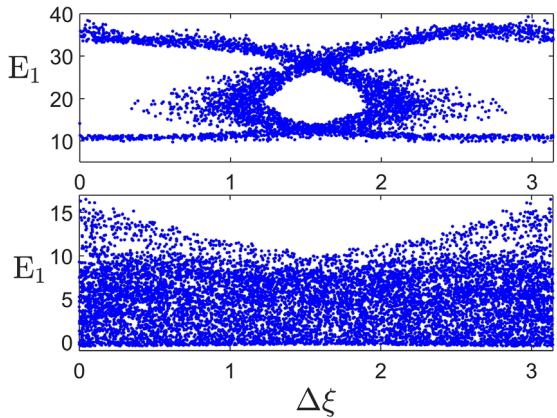

FIG. 3. The "map" obtained from numerical simulation for *case 1*, $a_0 = 1$, and initial energy $E_1^{(0)} = 4$ in the bottom panel and $E_1^{(0)} = 14$ in the top panel.

In Fig. 2, we show such maps corresponding to *case 1* for $a_0 = 0.5$ and $E_1^{(0)} = 4$ in the bottom panel and for $E_1^{(0)} = 14$ in the top panel. The stochastic region at high energy is bound by preserved KAM surfaces (see the top panel and the traces of the preserved "island" at high energy in the bottom panel). However, for higher $a_0$, the stochastic region extends to higher energies (see Fig. 3, which is obtained with the exact same conditions as Fig. 2 except for $a_0 = 1$ and higher initial energies), which is in agreement with the results of numerical simulations from Ref. 12. We notice that although initial electron "energy" is positive, with an onset of stochasticity, negative "energies" become accessible.

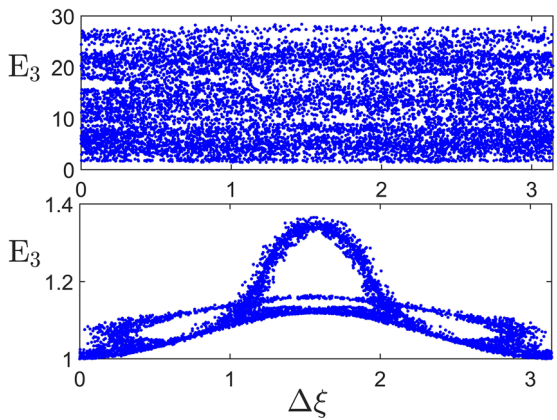

FIG. 4. The "map" obtained from numerical simulation for *case 3*, $a_0 = 0.5$, and initial energy $E_3^{(0)} = 1$ in the bottom panel and $E_3^{(0)} = 5$ in the top panel.

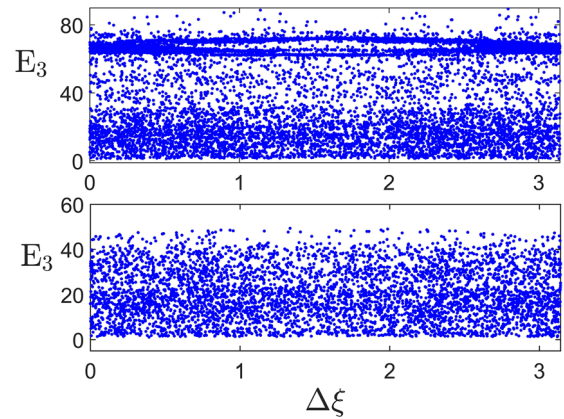

FIG. 5. The "map" obtained from numerical simulation for *case 3*, $a_0 = 1$, initial energy $E_3^{(0)} = 1$ in the bottom panel and $E_3^{(0)} = 5$ in the top panel.

In Fig. 4, we show such maps corresponding to *case 3* for $a_0 = 0.5$ and $E_3^{(0)} = 1$ in the bottom panel and for $E_3^{(0)} = 5$ in the top panel. For this set of initial data, the electron motion in some energy range is stochastic, while in other ranges, it is regular. However, an increase in the magnitude of $a_0$ to $a_0 = 1$ results in the "stochastization" of the wide energy range (see Fig. 5).

In conclusion, we analyze the interaction of electrons with intense linearly polarized laser radiation and static electric and magnetic fields in a 2D slab approximation, which is widely used for both numerical simulations and simplified semi-analytical models. We find that in this case, electron dynamics can be conveniently described in the framework of the 3/2 dimensional Hamiltonian approach. Based on this finding, we demonstrate that the electron acceleration beyond a standard ponderomotive scaling, caused by the synergistic effects of the laser and static electro-magnetic fields, is due to an onset of a stochastic electron motion (the transition to the stochastic regime occurs at $a_0 \sim 1$ for the parameters of static fields considered in this paper).

This work was supported by the University of California Office of the President Lab Fee Grant No. LFR-17-449059.